# Lock-in and Its Influence on the Project Performance of Large-Scale Transportation Infrastructure Projects

Investigating the Way in Which Lock-in Can Emerge and Affect Cost Overruns

By

Chantal C. Cantarelli, Bent Flybjerg, Bert van Wee, and Eric J. E. Molin

**Abstract**

Lock-in, the escalating commitment of decision-makers to an ineffective course of action, has the potential to explain the large cost overruns in large scale transportation infrastructure projects. Lock-in can occur both at the decision-making level (before the decision to build) and at the project level (after the decision to build) and can influence the extent of overruns in two ways. The first involves the "methodology" of calculating cost overruns according to the "formal decision to build". Due to lock-in, however, the "real decision to build" is made much earlier in the decision-making process and the costs estimated at that stage are often much lower than those that are estimated at a later stage in the decision-making process, thus increasing cost overruns. The second way that lock-in can affect cost overruns is through "practice". Although decisions about the project (design and implementation) need to be made, lock-in can lead to inefficient decisions that involve higher costs. Sunk costs (in terms of both time and money), the need for justification, escalating commitment, and inflexibility and the closure of alternatives are indicators of lock-in. In this paper, two case studies, of the Betuweroute and the HSL-South projects in the Netherlands, demonstrate the presence of lock-in and its influence on the extent of cost overruns at both the decision-making and project levels. This suggests that recognition of lock-in as an explanation for cost overruns significantly contributes to the understanding of the inadequate planning process of projects and allows development of more appropriate means.

# Introduction

Large-scale transportation infrastructure projects are often characterized by large cost overruns (Flyvbjerg et al, 2003; Hall, 1980; van Wee, 2007). Flyvbjerg et al. (2003) indicate that cost overruns appear in 86% of the projects under consideration with an average cost overrun of 28% (see Flyvbjerg (2005) for the definition of cost overrun).

Research has provided a variety of explanations (Flyvbjerg et al., 2002; Flyvbjerg et al., 2003; Hall, 1980; Pickrell, 1992; Wachs, 1989), which can be broadly grouped into four categories: technical, economic, psychological, and political (Flyvbjerg et al., 2003). The first, technical explanations involve "forecasting errors" expressed in technical terms; examples include imperfect forecasting techniques, inadequate data, and the lack of experience. The second, economic explanations involve issues of either economic self-interest or public interest. The third, psychological explanations include the concepts of planning fallacy (the tendency to underestimate the time needed to complete certain tasks) and optimism bias (the systematic tendency to be overly optimistic about the outcomes of actions), and the fourth, political explanations involve strategic misrepresentation, i.e. the deliberate and strategic underestimation of costs when forecasting the outcomes of projects.

Although researchers generally agree that the problem exists (van Wee, 2007), they differ widely about their causes and explanations. For example, both Flyvbjerg et al. (2003) and Pickrell (1992) consider strategic misrepresentation the main explanation, whereas Wachs (1989) explains the phenomenon in terms of ethical considerations. Because of this diversity in explanations, a broader view that addresses the phenomenon by including inadequate project planning in general is useful.

Arthur (1989), for example, uses a dynamic approach of allocation of resources under increasing returns to explain the outcome of a decision-making process. Based on the four principles of increasing returns, he explains how decision-makers select an outcome. First, the non-ergodic principle implies that different historical or chance events determine the drive towards a different outcome. Second, non-predictability indicates that the outcome could not be predicted before the historical or chance event took place. Third, the inflexibility principle concerns the lack of possibilities to influence the drive towards another outcome. Finally, path inefficiency indicates the presence of an outcome that would have paid off better. Policy results depend on the start and the specific development of the decision-making process in time (path dependency).

4Lock-in is created when sub-optimal policies are used as a consequence of path dependency, even though a better alternative is present (Woerdman, 2004). The term refers to the over-commitment of decision-makers to an ineffective course of action (e.g. a decision or project). Over-commitment itself refers to the style of psychological coping associated with the inability to withdraw from obligations (Vrijkotte et al, 2004). There are several possible moments in the decision-making process before the formal decision is taken at which decision-makers are committed to the project. This early commitment is, in itself, not necessarily negative and could also be advantageous to the decision-making process as it could enforce a decision, thus limiting delay. Early commitment can result in negative outcomes once the commitment turns into escalating commitment and lock-in. In this paper, as lock-in is based on escalating, it has, by definition, a negative influence on project performance. On this basis, this paper provides a more thorough examination of lock-in and its influence on the project performance, specifically regarding cost overruns, while considering the relationship between lock-in and the four categories of explanations mentioned above.

Although lock-in often concerns technical lock-in, (e.g. Paul David's article (1985) on the QWERTY keyboard) this paper is concerned with its institutional or behavioural form. It is a general phenomenon widely acknowledged in literature (e.g., Brockner et al, 1986; Staw, 1981; Whyte, 1986;), where the process of escalating commitment is also known as "entrapment" (Brockner and Rubin, 1985), the "sunk-cost effect" (Northcraft and Wolf, 1984), the "knee-deep-in-the-big-muddy" effect (Staw, 1976), and the "too-much-invested-to-quit" effect (Teger, 1980 in Brockner et al, 1986). However, institutional lock-in has never, to the best of our knowledge, been examined in the specific context of large-scale transportation-infrastructure projects.

In order to consider it, it is first necessary to understand how lock-in can influence the project performance. It can do so in two different ways. First, it can influence the extent through the "methodology" of calculating cost overruns. In such a case, a particular moment in time is often used to represent the moment at which the decision to implement the project was made ("formal decision to build") (Flyvbjerg et al, 2003). Cost overruns are commonly calculated according to the costs estimated at this "formal decision to build" (the costs at the initial funding level) point (Flyvbjerg et al., 2003). The decision-making process, however, involves several moments at which decisions are made so references to the formal "decision to build" do not always provide an accurate picture of cost overruns. In some cases, parties commit themselves at an earlier decision-making moment, known as the "real decision to build". In this paper, this situation is referred to as *lock-in at the decision-making level*. The reason why lock-in influences the extent of cost overruns, is because the estimated costs at the real decision to build are usually lower than those at later stages in the decision-making process.

In this paper, a distinction is made between the "formal" and "real" decision to build. In line with previous research in this field, a specific definition is used to determine the "formal decision to build", i.e. the moment at which the decision was taken in Parliament. Literature on this subject (see e.g. Flyvbjerg et al., 2003; Teisman, 1993) recognise that it is more difficult to determine the "real decision to build" because that decision is taken informally and within a fuzzy environment.

The second way in which lock-in can influence the extent of cost overruns is through "practice". Although the decision to implement the project has been made, specific decisions about the project itself also need to be made. These may not be 'optimal', may involve the danger of inefficient outcomes and can lead to *lock-in at the project level*.

To conclude, lock-in can thus be distinguished at two different levels, the decision-making and the project levels, and can influence the extent of cost overruns in two different ways, methodology and practice. In light of this, the general definition of lock-in is adjusted in this paper and formulated as: "the over-commitment of parties to an inefficient project before the formal decision to build and to the inefficient specifications of the project after the formal decision to build has been made."

The primary aim of this paper is to present lock-in within a framework in order to provide insight into the way in which it can actually occur and influence project performance (cost overruns). A further aim is to empirically determine whether lock-in has actually taken place in a project and, if it has, whether it has influenced the performance of the project. The main research question in this paper is, therefore, as follows: "Can lock-in provide an appropriate explanation for cost overruns?" This is answered by addressing two sub-questions: (1) "How can lock-in emerge at the decision-making and project levels?" and (2) "Has lock-in actually taken place in projects, and if so, how did it occur and until what moment in the decision-making process could the decision be reversed?"

Two research methods are applied to address these questions. First, a literature survey was conducted to address the first research question, and, second, case study research was used to derive empirical evidence to answer the second. The case-studies, HSL-South and Betuweroute projects, were specifically developed to determine whether lock-in was present in the decision-making process through the identification of several indicators, and, if it was, how it appeared. The projects were chosen because they are large-scale, well-documented and either nearly complete or recently implemented. In addition, since this study is part of a larger investigation of large-scale projects in the Netherlands, they are both Dutch projects.

Developments that took place in the project are examined and how these developments relate to the indicators of lock-in is investigated. In case many indicators of lock-in were present, it must be concluded that lock-in played a significant role in the decision-making process.

Since the reports of the Temporary Committee for Infrastructure Projects (TCI), which conducted extensive investigation of both projects, provide a good overview of their characteristics, they were used to derive most of the data. In addition, the semi-annual progress reports of the projects and several other reports were also consulted.

The scientific relevance of this work is that it fills the gap in knowledge concerning the contribution of lock-in to inadequate transport planning in large-scale transportation infrastructure projects. It is also of social relevance because of the major impact of lock-in on social welfare. Obtaining better insight in the role of lock-in in cost overruns is likely to lead to solutions that avoid it thus reducing the large burden to the state's budget.

This paper is organized as follows. Section 2 elaborates on the concept of lock-in and investigates how it can emerge and influence cost overruns. Sections 3 and 4 describe the two case studies, the Betuweroute and the HSL-South projects, respectively. Finally, section 5 presents the main conclusions and recommendations.

**Recognising lock-in**

This section addresses the first sub-question: "How can lock-in emerge at the decision-making and project levels?"

In order to define the ways in which lock-in can emerge, different indicators and criteria that determine whether the indicator is present are specified. Since, as the previous section indicated, path dependency plays a role in lock-in, two indicators were derived, inflexibility and closure of alternatives. Decision-makers who make a certain decision within an inflexible or incomplete (in the sense of not including all the alternatives) decision-making process are likely to be influenced by lock-in. A decision route becomes path dependency when previous decisions or events subject to inflexibility or closure of alternatives determine the current decision and the decision cannot be revised to reach another outcome.

Several studies have investigated the indicators for lock-in using various theories, including transaction costs economics (Amess, 2002), complexity theory (Walby, 2003), self-justification theory and prospect theory (Brockner et al, 1981; Brockner et al, 1982; Wilson and Zhang, 1997), and decision-dilemma theory (Bowen, 1987; Brockner et al, 1982). Transaction cost economics identifies *sunk costs* (i.e., irretrievable costs in terms of money and time) as an important subject of study: "Individuals show a greater tendency to continue an endeavour once an investment in funds, effort or time, has been made" (Mepham, 1987 in Brockner et al., 1982). As a consequence, sunk costs lead directly to lock-in at the project level and, through their impact on escalating commitment, indicate to lock-in at the decision-making level. The commitment to the project or decision increases concurrently with the amount of time invested in the decision-making process, making it more difficult to reconsider the decision.

The relationship between sunk costs and lock-in can also be explained by prospect theory, which describes how people make choices in situations in which they must decide between alternatives that involve risk. Kahneman and Lovallo (1993), argued that there is asymmetry in the way in which individuals value gains and losses; losses having a greater impact than equal gains. This asymmetry is known as loss aversion: the tendency to have a strong preference for avoiding losses over acquiring gains. Loss aversion can explain the sunk-cost effect as follows: when an investment in time or money is made (e.g., time spent in the decision-making phase or money spent in the project phase), individuals prefer to continue with the project because doing so allows for a chance of successful implementation, as opposed to a sure loss of the investment should they decide to quit. A decision is subject to sunk costs when a decision is taken to proceed with the project despite a lack of results from investments made in time or money.

Decision-makers show evidence of entrapment whenever they escalate their commitment to ineffective policies, products, services or strategies in order to justify previous allocations of resources to those objectives (Brockner et al, 1986). Escalating commitment and justification are therefore important indicators of lock-in. The *need for justification* is derived from the theories of self-justification and the theory of dissonance which describe how individuals search for confirmation of their rational behaviour (Staw, 1981; Wilson and Zhang, 1997). This need arises due to social pressures and "face-saving" mechanisms. The involvement of interest groups and organizational pushes and pulls can also introduce pressures into the decision-making process, threatening the position of the decision-makers, who may feel pressure to continue with a (failing) project in order to avoid publicly admitting

8what they may see as a personal failure (McElhinney, 2005). "People try to rationalize their actions or psychologically defend themselves against an apparent error in judgment" (Whyte, 1986) ("face-saving"). When the support for the decision is sustained despite contradicting information and social pressures, the argumentation for a decision is based on the need for justification.

Although *escalating commitment* can be the result of sunk costs and the need for justification, complexity theory and decision-dilemma theory have identified several other variables that lead directly to escalating commitment. Examples include agreements between parties or flawed decision-making processes that focus on solutions rather than on problems. Political vulnerability leads to escalating commitment in a project because of the influence of interests on decisions and commitments to decisions (Whyte, 1986). If strategic behaviour is evident in a project, it can also lead to escalating commitment, with individuals or groups acting in favour of a specific project, or underestimating the costs to make the project look better.

To determine if the argumentation of a decision was based on escalating commitment, the following indicators were identified: 1. an excessive focus on one outcome, 2. if agreements regarding the outcome were made previously, 3. strategic behaviour, or 4. actions were taken for political reasons.

These different theories identify several indicators of lock-in and provide insight into how the indicators are created and lock-in can result at the decision-making or project level. The relationship between the indicators and their outcomes is illustrated in the framework in figure 1, which shows the four indicators: sunk costs, the need for justification, escalating commitment, and inflexibility and the closure of alternatives.

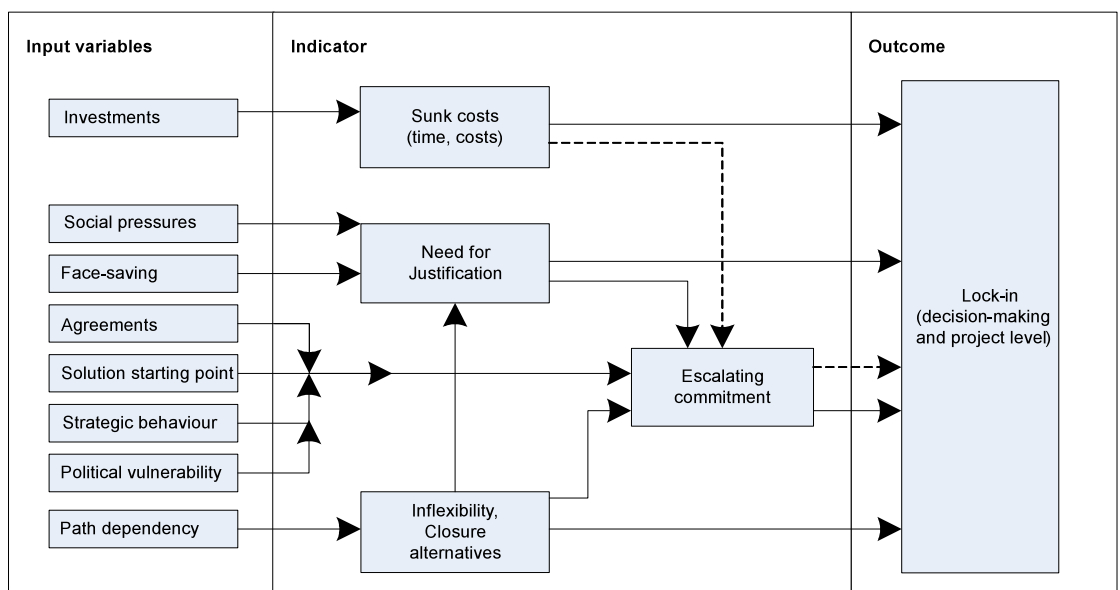

**Figure 1.** Theoretical framework for lock-in

A distinction should be made here between conscious and unconscious lock-in. With the former, decision-makers are aware of their tendency to justify decisions instead of evaluating them critically and have the possibility of reversing their decision. Unconscious lock-in, on the other hand, can occur in projects when the decision-maker cannot see a possibility of changing the situation. In Figure 1, the dotted lines represent the influence of unconscious lock-in, and the solid lines represent the influence of conscious. Sunk costs in terms of time lead to unconscious lock-in; escalating commitment, inflexibility and closure of alternatives lead to conscious lock-in. Since it is difficult to control sunk costs in terms of time, their impact is assumed to lead to unconscious lock-in. On the other hand, as other indicators can, to some extent, be managed, they are considered indicators of conscious lock-in.

There is a further distinction between intentional and unintentional lock-in. Lock-in can be a result of intentional behaviour on part of a decision-maker to ensure the implementation of a project. The importance of such lock-in is presented by Walby (2003), who argues that an important role is played by social and political institutions that lock-in certain paths of development, by shaping power, opportunity, and knowledge (Arthur, 1989; David, 1985; Mahoney, 2000; Nee and Cao, 1999; North, 1990; Pierson, 2000a; 2000b; 2001).

Some of the different types of lock-in can be avoided. For example, avoidance is when decision-makers are conscious about their behaviour and when they are willing to change their behaviour. However, other types of lock-in are more difficult to control, e.g. intentional lock-in which is a result of specific behaviour of parties.

Against the background of this insight, two case studies are described in section 3 and 4 with the aim of empirically considering the role of lock-in and providing an answer to the second sub-question: "Has lock-in actually taken place in projects and, if so, how did it occur and until what moment in the decision-making process could the decision be reversed?"

## Case study: The Betuweroute

The Betuweroute is a freight-transport railway line of about 160 kilometres, between the Port of Rotterdam and the European hinterland. It was finally opened in 2007 after a long decision-making process. As early as the early 1980s, the construction of a new railway line was proposed to deal with the unsatisfactory rail connections of the 'mainport' Rotterdam (Priemus, 2007), thus improving the connections with the hinterland and the mainport and strengthening the national economy (PKB part 1, design track decision). The Betuweroute project was approved by the Lower Chamber in 1994, after much public debate questioning its desirability and necessity. The project was then reconsidered but this did not influence the outcome, and, in 1996, the decision was taken to build the project. This case study involves a systematic search for the presence or absence of lock-in indicators (i.e., sunk costs, a need for justification, escalating commitment, inflexibility, and the closure of alternatives).

**Decision-Making Level**

The project was incorporated into the policy plans in 1990 (SVVII: Second Transport and Structure Plan) as a solution to the problem of insufficient railway capacity for freight transport to accommodate expected future growth. This created an excessive focus on the Betuweroute itself and shifted attention away from the problems (i.e., the *solution was taken as a starting point*) (TCI, 2004). Politician Hermans, chairman of the commission on the Betuweroute, concluded that "the Betuweroute was put on the agenda before any research was carried out on the alternatives" (De Gelderlander (1995 in Roscam Abbing et al., 1999). Priemus (2007) reach similar conclusions: "the solution was decided upon at a very early stage of the process". This excessive focus indicates the presence of *escalating commitment* by politicians to the project in the decision-making process. The number of *agreements*, another criterion for escalating commitment, also contributed to over-commitment to the project. Examples of this include the Agreement of Warnemünde (with Germany about the connection to the German railway network) and agreements relating to the project's inclusion in the SVVII. These agreements formalised the decision to construct the railway line.

The problem analysis remained narrow, focusing on identifying opportunities to develop Rotterdam Harbour as a 'main port' (Priemus, 2007), with the result that the decision-making process was inflexible and incomplete (focusing solely on railway connections instead of other options to increase the strength of the main port) and alternatives to the Betuweroute were not really considered.

Although the Betuweroute project was labelled as indicative rather than decisive in SVVII, the Dutch Railway company had investigated the specific implementation of the Betuweroute (TCI, 2004; Ministry of Transport, 1996-2007; Pestman, 2001) in the project Memorandum, thus limiting itself to studying this solution alone and failing to consider alternatives (e.g., zero plus alternatives, alternatives in modality, and rail alternatives). For example, a possible alternative, increasing the capacity of inland waterway transport, was never fully considered. Several other alternatives, e.g. joint hinterland connection by rail between Rotterdam and Antwerp, using the existing railway network more intensively and underground construction, were also not taken seriously or were proposed far too late to add anything to the discussion (Priemus, 2007).

Roscam Abbing et al. (1999) showed that reports were "written under order in which some comparisons were made and others were not, and for a government that was not willing to enter into any discussion concerning content and to stake everything to force a 'point' of no return". They argued that the question of how inland shipping might provide a possible solution was not answered as it could have harmed political support for the Betuweline. This shows escalating commitment as a result of the closure of alternatives and due to political vulnerability.

Social pressures as a consequence of continuous criticism of the project led to the Cabinet decision to follow a two-track policy, calling for the publication of the PKB1 while conducting additional research into the justification of the intention to implement the Betuweroute. This allowed them to continue with the Betuweroute project (further locking themselves into the project) while simultaneously addressing the criticism of the PKB1. Although the PKB-procedure had started, the desirability and necessity of the Betuweroute were yet to be discussed.

The need for justification can also be seen in the decision of the Cabinet to start a Key-Planning Decision (PKB) procedure as part of the New Track Law. Although this planning procedure provided a new opportunity to fundamentally question the project, steps that could not be reversed (lock-in) had already been taken. The TCI described the Cabinet's lock-in to the project as follows:

> "The Cabinet decided to take the lead and follow a PKB-procedure for the Betuweroute. This made it harder to reverse the decision for the project, which was exactly as the Cabinet had intended. The Cabinet used the time argument to pressure the parliament; postponing the decision-making was not desired".

Two new pieces of legislation (the New Track and NIMBY laws) were planned to prevent further delay to projects or, in other words, to speed up the decision-making process (Priemus and Visser, 1995). Overall, the reaction to the criticism of the Betuweroute led to face-saving and was the immediate cause for continuing with the project. This can also be seen as escalating commitment due to sunk costs in time.

Arguments to support the decision to implement the project were poorly founded. The report by the Hermans Commission (established in 1994 to investigate the desirability and necessity) lacked an adequately justified conclusion and the Nijffer research institute report provided no new insights into the economic effects of the project. Furthermore, the fact that the environmental objectives had not been met proved no reason for abandoning or adjusting the decision to implement the Betuweroute project (TCI, 2004; Ministry of Transport, 1996-2007). This decision shows face-saving mechanism on the part of the decision-makers. One of the main conclusions of Pestman's (2001) research into the Betuweroute was the importance of the need for justification:

> "whenever the degree of mobilisation is quite high, new insights and more information, which might prove advantageous in making a fair assessment of societal costs and benefits, are found. However, due to the mobilisation process, it becomes increasingly difficult for politicians to change their opinions because of their fear of losing credibility".

Even after environmental and economic arguments removed the foundations of the implementation decision, politicians insisted that it was a "strategic decision" (Abbing, 1999). This *political vulnerability* indicates the escalating commitment to the project created lock-in of the decision-makers to the project.

Finally, the unwillingness of political parties to change their opinion regarding the decision to implement the Betuweroute project once new insights into the desirability and necessity of the project became available indicates the presence of *inflexibility* in the project implementation. For example, the Netherlands Bureau for Economic Policy Analysis (CPB) concluded that the capacity along the East-West corridor was higher than expected, allowing for phased implementation. However, the decision-makers did not take this new information into account (inflexibility) (TCI, 2004 and Priemus 2007). This unwillingness to re-open the discussion again created lock-in through the failure to consider other, potentially more efficient, implementation choices.

Figure 2 (based on TCI, 2004; Ministry of Transport, 1996-2007) shows the decision-making moments that led to lock-in along a timeline:

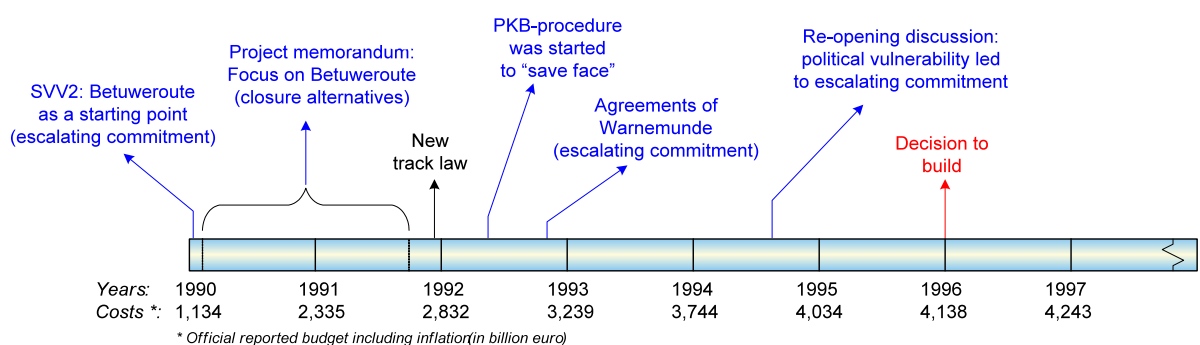

**Figure 2.** Timeline for the Betuweroute project (decision-making level)

Although the formal decision to build was taken in 1996, all five decision-making moments had taken place before 1996, suggesting the presence of lock-in in the decision-making process. It is therefore reasonable to assume that the real decision to build was made before 1996. Identifying the "real decision to build" requires the consideration of each decision-making moment with regard to the relative possibility of revising the decision. The moment at which this was no longer possible is known as the "point of no return". In 1990, when the SVVII was published, the project was described as "indicative" but no formal agreements had yet been made. Consequently, it was still possible to reconsider the decision. This was also the case for the decision-making moment of the project memorandum. Although the memorandum considered only the Betuweroute and no further alternatives, it did not incorporate any agreements that precluded the possibility of reversing the decision. The "point of no return" in the Betuweroute project was actually the start of the PKB-procedure in 1992, at the start of which the first steps into project implementation had already been taken even though a clear justification was lacking,. This made it impossible to re-open the discussion concerning the decision to build. Withdrawing was no longer an option, decision-makers were obliged to the project.

Lock-in thus led to a real decision to build that was taken before the formal decision to build. Consequently, the extent of cost overrun at this real decision to build also differs. Cost estimates earlier in the decision-making moment are usually much lower and the extent of cost overrun is therefore expected to be higher at the real decision to build. More specifically, cost overruns calculated in connection with the real decision to build are 64.6% (with final costs 2007 of 4.663 billion euro and base line funding year 1992) while cost overruns calculated in connection with the formal decision to build (with final costs 2007 of 4.663 billion euro and base line funding year 1996) were "merely" 12.7%. Lock-in thus resulted in the inaccurate representation of the extent of cost overruns. Note that the percentage cost overrun calculated at the "formal decision to build" in this paper differs from figures presented in former studies. This can be explained by the status of the project at the time of the study. In previous research, the project was not yet complete and the budget for the total project when 88% of this budget had been spent was therefore used as the 'actual costs'.

**Project Level**

An indication of lock-in at the project level is the need for justification by decision-makers as a consequence of the new budget-control philosophy of "steering on a limited budget". This philosophy is applied in situations in which the actual costs prove higher than expected, thus raising the threat of a deficit. In order to deal with this, the budget can either be increased or the scope of the project adjusted. This philosophy ensures that, in such a case, the budget will not be re-adjusted but instead control measures will be applied to ensure that the problems are resolved within the limited budget.

Furthermore, lock-in is seen in the escalating commitment to the project despite financial tensions during project implementation.

The "Malle Jan" arrangement (between the Minister and the project organization that called for the realization of the Betuweroute project within the specified requirements) had been established to resolve this tension but did not have the desired results and project implementation continued despite negative results, with the budget being changed at the expense of the project scope.

In addition to the *escalating commitment*, the project involved *closure of alternatives* (e.g. with respect to ground-level or underground construction). The provinces of Gelderland and Zuid-Holland proposed underground construction as an alternative. According to the van Engelshoven Steering Group, established to investigate the possibility of underground construction, this alternative was both too expensive and too risky. The Minister's firm standpoint on the type of construction created lock-in, and other construction methods (e.g., underground construction) failed to receive fair consideration.

The existence of lock-in at the project level was further indicated by the *inflexibility* regarding financing and the discussion concerning the desirability and necessity of the project. An assessment scheme to determine the desirability and necessity of the project was also poorly founded (Pestman, 2001).

With regard to financing, private financing was adopted as the starting point. This option ultimately proved impossible but by then the project was already at a stage at which abandonment was no longer an option. Due to inflexibility with regard to changing the decision, political parties were locked-in to the project. Some parties felt deceived, as they had considered private financing a precondition for the implementation decision. The starting assumption of private financing created lock-in, and may have precluded the examination of other types of potentially more appropriate financing.

On the other hand, there were some decisions that did include flexibility at the project level. For example, the decision regarding the crossing of the Pannerdensch Kanaal left room for discussion about whether it was to be by a bridge or via a tunnel (Pestman, 2001). In terms of the construction method of the tunnels, flexibility was also seen. In the end, the tunnels were bored, a situation not foreseen at the start of the decision-making process.

Figure 3 shows the decision-making moments that led to lock-in. Since the "Malle Jan" agreement was established solely to control costs, it had no direct impact on the cost overruns. The new philosophy did, however, influence project performance as, given that construction had not yet begun. It was still possible to withdraw from the project at that time.

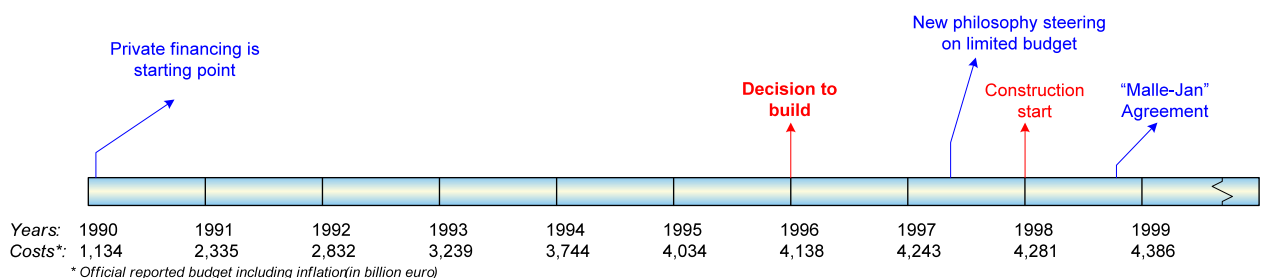

**Figure 3.** Timeline for the Betuweroute (project level)

## Case study: HSL-South

This case study involves a systematic search for the presence or absence of lock-in indicators at the decision-making and project levels. After the success of the high-speed railway connection between Paris and Lyon, the idea of a European network of high-speed railway trains emerged. In 1986, the ministers responsible in France, Belgium, Germany, and the Netherlands agreed to develop an HSL-network between Paris, Brussels, and Amsterdam, with the HSL-South as the Dutch portion. Procedures of a PKB (key planning decision) started in 1986 and, after several delays, part of the HSL-South (the part between Amsterdam and Rotterdam) opened in September 2009.

**Decision-Making Level**

In the decision-making process of the HSL-South, path dependency played an important role. During international consultations, decisions were taken about the mode of transport, time schedule, financing and the specific characteristics of the different tracks (de Vries et al., 2007). For example, a decision was taken not to use existing infrastructure for the main part of the railway, but to construct new infrastructure. With regard to the design speed, decisions had also already been taken and the railway line had to be suitable for high-speed traffic of 300 km/h. These decisions at the international level set limitations on the decisions to be made at the national level, creating inflexibility for national policy. This was seen in the SVVII (Second Transport and Structure Plan), which confirmed connection of the Netherlands to the European network of high-speed railway lines and the construction of a new railway line between Rotterdam and the Belgium border suitable for high speeds of 300 km/h. With the acceptance of this plan by the Lower Chamber, the Netherlands embedded international agreements in national policy (de Vries et al., 2007). These agreements, which made decisions binding, indicate the over-commitment to the project. Furthermore, there was escalating commitment in the assumptions regarding the HSL-South project: despite the lack of any conclusions about the desirability or necessity of the high-speed railway line, the government considered the Dutch connection to the European HSL-network essential. This illustrates a common problem in the decision-making surrounding large projects: "the solution, rather than the problem, was taken as a starting point". The geographical location and the advantages regarding economy, transport value and the environment were afterwards advanced as justification for the project. Most of the discussion on the HSL-South was related to the track decision. Though the different possibilities had been determined by previous decisions, the Netherlands and Belgium preferred different options and the focus was on these two preferred options. Agreement was eventually reached as a consequence of political vulnerability: the minister wanted to publish the new plan (SVVII) while she was still in office (escalating commitment to the plan).

Lock-in at the decision-making level was created by the *inflexibility* of the decision-making process regarding deviations from prior decisions. In the formation of a new Cabinet, the Ministry of Transport supported the decisions already made by referring to the coalition agreement: "the decision to implement the high-speed railway line is confirmed, including the track choice." It was argued that any deviations from these decisions would lead to practical objections and negative effects on the economic position of the Netherlands. Different alternatives were included in the consultation round but did not stand a fair chance of success (Priemus, 2007).

The presence of lock-in in the decision-making process of the HSL-South is described by Priemus (2007) as follows:

> "By rigidly maintaining the design speed requirement of 300 km/h, and by including the preferred alternative in the 1994 Coalition Agreement, the cabinet was able to pass this choice through the political and social decision-making procedures".

There were, however, also several occasions in the decision-making process that contradict the presence of escalating commitment, inflexibility or the closure of alternatives. The first plan (SVVI) was received poorly, leading to a new plan being composed. This possibility and the willingness to change the plan point to flexibility in the decision-making process. There was also flexibility in the decision-making process regarding the stops: although not originally a stop, Antwerp was, it was eventually included due to social pressure. The decision-making moments that led to lock-in are presented in Figure 4.

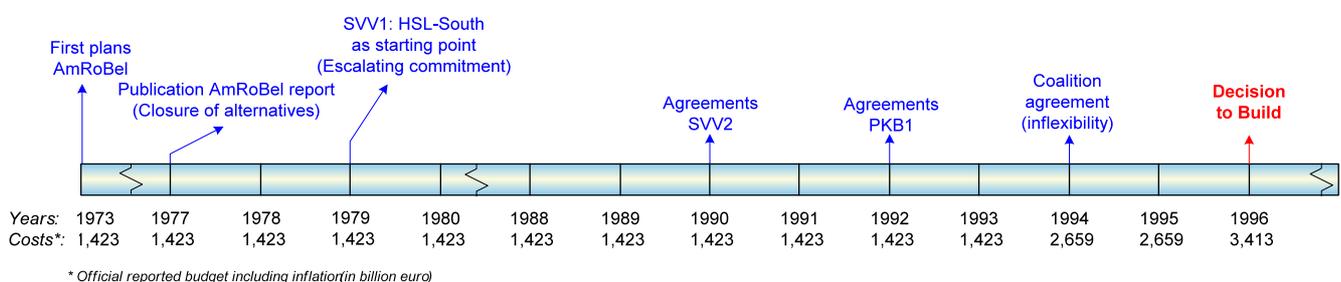

**Figure 4.** Timeline of the HSL-South project (decision-making level)

The decision-making moments identified preceded the formal decision to build (which occurred in 1996). The real decision to build was, therefore, probably made before the formal decision. Once again, identifying the real decision to build is likely to help determine the influence of lock-in on project performance in terms of money. In other words, the point of no return must be established by considering the possibility of reversing the decision at each of the decision-making moments. At decision-making moments SVVI and SVVII, it was still considered possible to reconsider the decision to build, as no agreements had actually been made. As in the Betuweroute project, the actual point of no return in the HSL-South project was marked by the start of the PKB-procedure.

The extent of cost overruns are even higher if the PKB1 decision-making moment in 1992 is taken as a reference instead of the formal decision to build in 1996 (403.67% as compared to 110%, with final costs of 7.17 billion euro). Note again that these percentages may differ from other studies as a consequence of the status of the project. Cost overruns based on the 'formal decision to build' are therefore misleading and lock-in actually resulted in even higher cost overruns.

**Project Level**

*Sunk costs* also created lock-in at the project level with the difference that project-level sunk costs involved money and those at the decision-making level involved time. Investments led to unwillingness to abandon the project after the formal decision to build had been made, thus creating lock-in with regard to the project.

The threat of cost overruns created a need for a new budget-control method, leading to the establishment of the philosophy of "steering on a limited budget". There was an excessive focus on the price, however, at the expense of the risks, scope, design, and quality of the contracts to be made. Despite negative results with the philosophy (e.g. social pressures and face-saving behaviour by parties who felt unable to admit their mistakes), it was decided to proceed with it, indicating lock-in with regard to the budget-control measure. In addition, problems related to the budget led to strategic behaviour on the part of contractors. The low tender budgets and estimations were the immediate cause of the structural underestimation of investment costs in the five contracts. Furthermore, the need for justification is evident from the discussion on private financing. The project had originally been based on this, and assumed a contribution of fifty percent. This approach was, however, apparently based more on departmental ambition than on the demands of market parties. The need to justify the assumed contribution of private financing therefore created lock-in with regard to the type of financing.

The *closure of alternatives* regarding track choice and design speed provides further evidence of lock-in. Although the Cabinet and the House of Representatives had reached an impasse with regard to track choice, the situation had little influence on the actual track choice decision. The Cabinet's preferred track choice (A1, a new track on the east side of Zoetermeer, crossing the Groene Hart) was implemented. The Cabinet specified a design speed of 300 km/h. The closure of alternatives for track choice and design speed was evident in the Cabinet's insistence on sticking to these decisions in spite of the lack of clear justification. In the end, none of the arguments to support the decision of the high design speed proved to be accurate. Thus, the Cabinet's insistence created lock-in with regard to track choice and design speed, precluding the possibility of considering other tracks and design speeds that may have been more efficient. Finally, the detailed reference design provided to the building contractors as a basis for submitting the tender indicated *inflexibility* in the project. The limited amount of design freedom created lock-in to the reference design, thus hampering the any search for more efficient designs.

Figure 5 shows the decision-making moments that led to lock-in. As most of the decision-making moments that led to lock-in were created before actual construction started, reversing lock-in and reconsidering the decision to build was still possible.

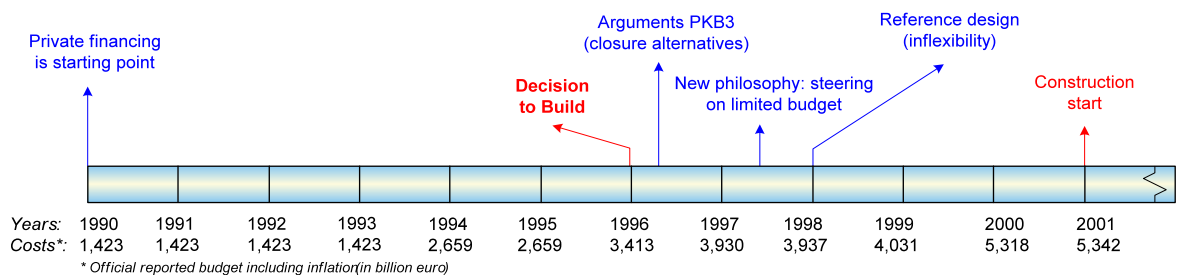

**Figure 5.** Timeline of the HSL-South (project level)

## Conclusions and recommendations

On the basis of this research into how lock-in can occur and how it can influence cost overruns, the main conclusions are as follows: lock-in can appear at both the decision-making and the project levels, and it can lead to cost overruns through methodology and practice. The presence of lock-in in the decision-making process and in the project can be demonstrated by the presence of sunk costs, escalating commitment, need for justification, inflexibility, and the closure of alternatives. In both the case studies, lock-in occurred at both the decision-making and project levels. At the decision-making level, the actual decision to build the projects preceded the formal decision to build. More specifically, although the formal decision to build for both projects was made in 1996, the point of no return was passed previously, in 1992. This moment was marked by the beginning of the PKB procedure, which can be seen as the real decision to build. The difference between the formal and real decisions to build indicates lock-in, which influenced cost overruns through methodology. The costs estimated at the real decision to build were much lower, thus increasing the actual cost overruns. At the project level, higher costs resulted from limited freedom to change possibly inefficient decisions regarding the design of the project.

The extent and appearance of lock-in is different in the two different case studies. The Betuweroute project was widely debated public and faced a lot of criticism, resulting in decisions driven by the need for justification and escalating commitment. The HSL-South, on the other hand, had considerable support at the start of the project and the need for justification was therefore less dominant. In both cases, lock-in emerged at a very early stage because one of the main criteria for escalating commitment, the solution, was taken as a starting point. If decision-makers had been aware of this, lock-in could have been limited to a certain extent. However, the danger of lock-in would have remained during the whole decision-making and project phases. In this paper, a distinction was made between early commitment, escalating commitment and lock-in. Lock-in is by definition a negative phenomenon. When the scope is enlarged, early commitment could be advantageous, e.g. regarding early land development and shorter procedures.

Lock-in has two important *policy implications*. First, lock-in by methodology necessitates identification of the decision-making moments representing both the formal and the real decision to build in order to determine whether there is a difference and whether any methodological corrections are needed in the calculation of cost overruns. Second, with regard to policy implications for practice, cost overruns can be avoided if lock-in is prevented. Some decision-makers deliberately create over-commitment or exclude other alternatives in order to create lock-in for their preferred projects or decisions. The subsequent cost overruns may thus be partly unnecessary or avoidable due to the deliberate creation of the lock-in, and be partly unavoidable due to the incapacity of decision-makers to make optimal decisions (i.e. bounded rationality).

The findings on lock-in reported in this paper have further important *implications for theory*. To summarise, lock-in can actually be placed within each of the four categories, technical, economic, psychological and political explanations. In contrast Flyvbjerg et al. (2003), which argues that technical explanations are least likely to explain cost overruns, here the concept of lock-in has proved that technical explanations also constitute are an important category of explanations. Lock-in also stresses the importance of economic and psychological explanations, with sunk costs in terms of money creating conscious lock-in as various parties are aware of their investments, while the bounded rationality of decision-makers results in unconscious lock-in. Finally, lock-in is a psychological explanation if it arises from behaviour intended to justify decisions, and is a political explanation if it emerges in response to intentional (strategic) behaviour. This study shows that despite the significant theoretical impact that the methodology used in a project can have on its performance, the difference is more difficult to demonstrate in practice. Although it is known that earlier decision-making moments lead to higher cost overruns, the extent of the difference between the actual and formal decision-making moments remains to be proven. Subsequent research into lock-in is therefore recommended. Once interviews have been held with decision-makers in different projects and the actual decision-making moments established, the actual and formal decisions to build can be compared, allowing more precise conclusions concerning the influence of lock-in on project performance through the methodology adopted.

## Acknowledgement

This research was supported by the Dutch Ministry of Transport. The authors thank two anonymous referees for their useful comments. The authors thank Eddy Westerveld of AT Osborne for his valuable help.